\documentclass[12pt,a4]{article}

\usepackage{amsmath}
\usepackage{amssymb}
\usepackage[latin1]{inputenc}

\begin{document}

\renewcommand{\theequation}{\thesection.\arabic{equation}}
\thispagestyle{empty}
\vspace*{-1.5cm}

\begin{center}
{\large THE CONSTRUCTION OF TRIGONOMETRIC\\
INVARIANTS FOR WEYL GROUPS AND THE \\
DERIVATION OF CORRESPONDING EXACTLY \\
\vspace{0.12cm}
SOLVABLE SUTHERLAND MODELS }\\

\vspace{2 cm}
{\large O. Haschke and W. Rühl}\\
Department of Physics, University of Kaiserslautern, P.O.Box 3049\\
67653 Kaiserslautern, Germany \\
\vspace{5cm}
\begin{abstract}
Trigonometric invariants are defined for each Weyl group orbit on the root lattice. They are real and periodic on the coroot lattice. Their polynomial algebra is spanned by a basis which is calculated by means of an algorithm. The invariants of the basis can be used as coordinates in any cell of the coroot space and lead to an exactly solvable model of Sutherland type. We apply this construction to the $F_4$ case. 

\end{abstract}
\vspace{5cm}
{\it March 1999}
\end{center}
\newpage

\section{Introduction}
Integrable models of the Calogero-Moser class and their trigonometric and rational limit models are conventionally described by the simple Lie algebras, i.e. those contained in the classical sequences $A_n, B_n, C_n, D_n$ or the exceptional set $G_2, F_4, E_6, E_7, E_8$. We have shown \cite{1,2} that the Weyl groups underlying these algebras are the essential ingredients in the construction of the trigonometric or rational models. Exact solvability is easily  proven by expressing the Schrödinger equation in terms of the Weyl group invariants as coordinates. For the rational models the crystallographic property of the Weyl groups, which guarantees the existence of a root lattice, can be abandoned. For the proper Coxeter groups $H_3$ and $H_4$ and the infinite sequence of dihedral groups \cite{2,3} Calogero type models can be constructed as well.

For the classical Lie algebras and $G_2$ the Weyl group invariants are not derived but guessed by intuition \cite{4}. But this method failed for the exceptional algebras $F_4$ and $E_6, E_7, E_8$. Therefore a concept supplying us with all trigonometric polynomial invariants and their algebraic basis is highly desirable. In the case of the rational  polynomial invariants the existence of the algebraic basis is guaranteed by Chevalley's theorem \cite{5}. The Jacobian for the transition from cartesian coordinates in root space to the basic invariants as new coordinates can be factorized (``factorization theorem''). Both the Chevalley theorem and the factorization theorem are valid also in the trigonometric case as we shall show here for $F_4$ (for the classical Lie algebras and $G_2$ they are known to be valid, too).

Our construction of invariants proceeds as follows. We decompose the root lattice $\Lambda$ into orbits $\Omega$ (infinitely many, they can be ordered by the length of vectors which is constant over the orbit). For each orbit we define an invariant trigonometric polynomial
\begin{equation}
T_\Omega(x) = \sum_{\beta \in \Omega} \exp i (\beta, x)
\label{1.1}
\end{equation}
These functions obey fusion rules
\begin{equation}
T_{\Omega_1}(x) T_{\Omega_2}(x) = \sum_{\Omega_3} C^{\Omega_3}_{\Omega_1\Omega_2} T_{\Omega_3}(x)
\label{1.2}
\end{equation}
with ``fusion coefficients'' $C^{\Omega_3}_{\Omega_1\Omega_2}$ that are nonnegative integers. In (\ref{1.2}) the null-orbit consisting only of the null-vector in $\Lambda$ must be included 
\begin{equation}
T_{\Omega_0}(x) = 1
\label{1.3}
\end{equation}
The system of equations (\ref{1.2}) is of triangular shape and can be solved trivially for $T_{\Omega_{\max}}(x)$. Each pair $\Omega_1, \Omega_2$ defines in fact a unique $\Omega_{\max} (\Omega_1, \Omega_2)$ with
\begin{eqnarray}
C_{\Omega_1\Omega_2}^{\Omega_{\max}(\Omega_1,\Omega_2)} = 1
\label{1.4}
\end{eqnarray} 
By recursive substitutions we isolate then an algebraic basis
\begin{eqnarray}
& &\left\{ T_{\Omega_1}, T_{\Omega_2}, ... T_{\Omega_n} \right\} \\ \nonumber
& &n = \mbox{rank (Weyl group)} 
\label{1.5}
\end{eqnarray}
and for all other orbits we obtain explicitly
\begin{equation}
T_\Omega(x) = {\rm pol} \{T_{\Omega_1}(x),...T_{\Omega_n}(x), T_{\Omega_0}(x) \}
\label{1.6}
\end{equation}
This is an explicit and constructive version of Chevalley's theorem. It is obtained only in a case-by-case study (e.g. for $F_4$). In Section 2 we do this in great detail for $F_4$. 

In Section 3 we apply this technique to construct the $F_4$ Sutherland model, using the approach developed in \cite{1,2}. By the way some minor theorems are proven by explicit calculation (e.g. existence of the $r_i^{(a)}$-coefficients as polynomials in the Chevalley basis).

\setcounter{equation}{0}

\section{Trigonometric invariants of Weyl groups. }

Weyl groups are generated by reflections along roots $\alpha \in \mathbb{R}_n$ \cite{5}
\begin{equation}
x \in \mathbb{R}_n: s_\alpha x = x - 2 \frac{(\alpha,x)}{(\alpha,\alpha)} \alpha 
\label{2.1}
\end{equation}
Each root $\alpha$ is an integer linear combination of simple roots
\begin{equation}
\{ \alpha_1, \alpha_2,...\alpha_n \}
\label{2.2}
\end{equation}
These simple roots span an integral lattice $\Lambda \in \mathbb{R}_n$
\begin{equation}
\beta \in \Lambda: \; \beta = \sum^n_{i=1} m_i \alpha_i, \quad m_i \in \mathbb{Z}
\label{2.3}
\end{equation}
If the Weyl group $W$ acts on a vector $\beta \in \Lambda$ it produces an orbit $\Omega$
\begin{equation}
\Omega = \{ w \beta, \; w \in W \}
\label{2.4}
\end{equation}
How can such orbit be characterized?

Since $W$ is a discrete subgroup of $O(n)$ acting on $\mathbb{R}_n$ we obtain
\begin{eqnarray}
\| w \beta \|^2 &=& (w\beta, w\beta) \nonumber \\
&=& \| \beta \|^2
\label{2.5}
\end{eqnarray}
so that an orbit appears as  a discrete set on a sphere of radius $\| \beta \|$ and (see (\ref{2.3}))
\begin{equation}
\| \beta \|^2 = \sum_{i,j} (\alpha_i, \alpha_j) m_im_j
\label{2.6}
\end{equation}
Here $(\alpha_i, \alpha_j)$ is contained in the Cartan matrix as
\begin{equation}
A_{ij} = 2\frac{(\alpha_i,\alpha_j)}{\| \alpha_i \|^2}
\label{2.7}
\end{equation}
Only for simply laced Weyl groups the lengths of the simple roots are all equal.

For the non-simply laced Weyl group $W = F_4$ we use as basis in $\mathbb{R}_4$
\[ \{ e_i, \; i \in \{1,2,3,4\}, \quad (e_i,e_j) = \delta_{ij} \} \]
Then $W = F_4$ can be generated from the reflections along $e_i$ and $f_i$ (all $i$) where
\begin{eqnarray}
f_1 &=& \frac12 (e_1+e_2+e_3+e_4) \nonumber \\
f_2 &=& \frac12 (e_1 + e_2 - e_3 - e_4)  \nonumber \\
f_3 &=& \frac12 (e_1-e_2+e_3-e_4)  \nonumber \\
f_4 &=& \frac12 (e_1-e_2-e_3+e_4)
\label{2.8}
\end{eqnarray}
\begin{equation}
(f_i,f_j) = \delta_{ij}
\label{2.9}
\end{equation}
and the permutation group $S_4$ of the basis $\{ e_i \}$. The set of roots decomposes into two orbits
\begin{equation}
\Omega_1 = \{ \pm e_i, \; \frac12 (\pm e_1 \pm e_2 \pm e_3 \pm e_4) \}
\label{2.10}
\end{equation}
\begin{equation}
\Omega_2 = \{ \pm e_i \pm e_j, \; i < j \}
\label{2.11}
\end{equation}
which are characterized by a subscript denoting the length squared of the roots. In addition we need the null-orbit
\begin{equation}
\Omega_0 = \{ \mbox{null-vector} \}
\label{2.12}
\end{equation}
All other orbits can be characterized by an integral radius squared and a further ``degeneracy'' label. A list of orbits up to $\| \beta\|^2 = 24$ is given in Table 1.

Trigonometric invariants are defined for each orbit by
\begin{equation}
T_{n,a}(x) = \sum_{\beta \in \Omega_{n,a}} \exp i (\beta, x)
\label{2.13}
\end{equation}
(i.e. $\| \beta \|^2 = n$), so that
\begin{equation}
T_{n,a}(0) = \# \Omega_{n,a}
\label{2.14}
\end{equation}
Now we consider pairs of such trigonometric invariants and expand their product as
\begin{equation}
T_{n,a}(x) T_{m,b}(x) = \sum_{k,c} C^{(k,c)}_{(n,a)(m,b)} T_{k,c}(x)
\label{2.15}
\end{equation}

\noindent
Applying permutations of $S_4$, reflections along the coordinate axis $e_i, \, i \in \{ 1,2,3,4 \}$ and reflection along $f_1$ to any generating vector yields the whole orbit.

\vspace{1cm}
We expand $x \in \mathbb{R}_n$ in a co-root basis 
\begin{equation}
x = \sum_i \xi_i \tilde{\alpha}_i
\label{2.16}
\end{equation}
Then by inversion of (\ref{2.15}) we obtain
\begin{eqnarray}
(2\pi)^{-n} \int_{\rm cell} d^n \xi T_{m,a}(x) T_{n,b}(x) T_{k,c}(x) \nonumber \\
= \# \Omega_{k,c} \cdot C^{(k,c)}_{(m,a) (n,b)}
\label{2.17}
\end{eqnarray}
Note that the r.h.s. is symmetric in the three orbits. We denote (\ref{2.15}), (\ref{2.17}) the ``fusion rules'' for trigonometric invariants. The ``fusion coefficients'' $C^{(k,c)}_{(m,a),(n,b)}$ are nonnegative integers.

We expect that a Chevalley theorem of the following type is valid: \\
Any polynomial in the $\{T_{n,a}(x)\}$ can be expressed as a polynomial in an algebraic basis of invariants including $T_{\Omega_0} = 1$. The number of nontrivial basis elements is rank $W$.

In the case of $W = F_4$ this algebraic basis is constructed by inversion of (\ref{2.15}) and consists of 
\renewcommand{\thefootnote}{*)}
\begin{equation}
T_1, T_2, T_3, T_6 \footnote{}
\label{2.18}
\end{equation}
\renewcommand{\thefootnote}{}
\footnote{*) We could replace $T_6$ by one of the following invariants: $\{T_7, T_8, T_{9,1}, T_{9,2}, T_{10} \}$ or constant linear combinations thereof.}
This inversion is possible by the triangular shape of the fusion rules. Namely inserting (\ref{2.13}) into (\ref{2.15}) and using the triangular inequality we obtain
\begin{eqnarray}
C^{(k,c)}_{(n,a)(m,b)} = 0 & &  \mbox{except possibly for} \nonumber \\
& & |\sqrt{n} - \sqrt{m}| \le \sqrt{k} \le \sqrt{n} + \sqrt{m}
\label{2.19}
\end{eqnarray}
Thus there is a maximal $k$ for each $n,m$
\begin{equation}
k_{\max} \leqq (\sqrt{n} + \sqrt{m})^2
\label{2.20}
\end{equation}
and we solve (\ref{2.15}) for $T_{k_{\max},c}(x)$ where $c$ is such that the fusion coefficient is one. The result for
\begin{equation}
T_{k,c}, \quad k \notin \{ 1,2,3,6 \}, \quad k \leq 24
\label{2.21}
\end{equation}
expressed as a polynomial in $T_1, T_2, T_3, T_6$ is given in Table 2.

The task to introduce the trigonometric invariants as coordinates in each cell of the space $\mathbb{R}_n$ leads to the study of the Jacobian matrix
\begin{equation}
\left\{ \frac{\partial T_m}{\partial x_i} \right\}_{m \in \textrm{\scriptsize{basic set of invariants}}}
\label{2.22}
\end{equation}
In the program of constructing exactly solvable models we have to compute the (inverse) Riemannian
\begin{eqnarray}
g^{-1}_{mn} &=& \sum_i \frac{\partial T_m}{\partial x_i} \frac{\partial T_n}{\partial x_i} (x) \nonumber \\
&=& - \sum_{\beta \in \Omega_m} \sum_{\beta \in \Omega_n} (\beta,\beta^\prime) \exp i (\beta + \beta^\prime, x) \nonumber \\
&=& - \frac12 \sum_{k,a} (k-m-n) C^{(k,a)}_{(m)(n)} T_{k,a} (x)
\label{2.23}
\end{eqnarray}
which can obviously be expressed as a polynomial in the basic invariants. For $F_4$ the largest orbit appearing is $\Omega_{24}$ in $g^{-1}_{66}$.


\setcounter{equation}{0}

\section{The $F_4$ Sutherland model}

From (\ref{2.23}) a first version of the Riemannian is obtained by insertion of the fusion coefficients
\begin{eqnarray}
g^{-1}_{11} &=& -T_4 - T_3 + 4T_1 + 24 \nonumber \\
g^{-1}_{12} &=& -T_5 + 6T_1 \nonumber \\
g^{-1}_{13} &=& - \frac32 T_7 - 3T_6 - 2T_5 + 3T_3 + 12T_2 + 12T_1 \nonumber \\
g^{-1}_{16} &=& -2T_{11} - 2T_{9,1} + 4T_5 + 6T_3 \nonumber \\
g^{-1}_{22} &=& -2T_8 -2T_6 + 8T_2 + 48 \nonumber \\
g^{-1}_{23} &=& -2T_{9,1} - 3T_7 + 6T_3 + 24T_1 \nonumber \\
g^{-1}_{26} &=& -3T_{14} - 6T_{12} - 4T_{10} +6T_6 + 24T_4 + 24T_2 \nonumber \\
g^{-1}_{33} &=& -3 T_{12} -5 T_{11} -8 T_{10} -12 T_{9,2} -6 T_{9,1} -12 T_8 -3 T_7 +4 T_5  \nonumber \\
& & +12 T_4 +21 T_3 +48 T_2 +60 T_1 +288 \nonumber \\
g^{-1}_{36} &=& -4T_{17,1} - 6T_{15} - 6T_{13,2} - 8T_{13,1} -4T_{11} \nonumber \\
& & + 6T_7 + 16T_5 + 36T_3 + 48T_1 \nonumber \\
g^{-1}_{66} &=& - 6T_{24} -10T_{22} - 16T_{20} - 12T_{18,2} - 24T_{18,1} \nonumber \\
& & - 24T_{16} - 6T_{14}+ 8T_{10} + 24T_8 + 42T_6 +96 T_4 +120 T_2 +576 \label{3.1}
\end{eqnarray}
A second form is obtained by substitution of the algebraic basis $T_1,T_2,T_3,T_6$ by means of Table 2.
\begin{eqnarray}
g^{-1}_{11} &=& -T^2_1 + T_3 + 6T_2 + 12 T_1 + 48 \nonumber \\
g^{-1}_{12} &=& -T_1T_2 + 3T_3 + 12T_1 \nonumber \\
g^{-1}_{13} &=& - \frac32 T_1T_3 + 4T_1T_2 + 12T^2_1 + \frac32T_6 - 24T_3 - 42T_2 - 96T_1 - 288 \nonumber \\
g^{-1}_{16} &=& -2T_1T_6 + 2T_2T_3 + 4T_1T_2 - 12T_3 - 48T_1  \nonumber \\
g^{-1}_{22} &=&  -2T^2_2 + 12T^2_1 + 2T_6 - 24T_3 - 48T_2 - 96T_1 - 192 \nonumber \\
g^{-1}_{23} &=&  -2T_2T_3 + 3T_1T_3 - 4T_1T_2 - 24T^2_1 - 9T_6 + 60T_3 \nonumber \\
& & + 108T_2 + 240T_1 + 576 \nonumber
\end{eqnarray}
\begin{eqnarray}
g^{-1}_{26} &=&  -4T^2_1T_2 - 24T^3_1 - 3T_2T_6 + 3T^2_3 + 8T_3T_2 - 6T_6T_1 \nonumber \\
& & + 72T_3T_1 + 104T_2T_1 + 12T^2_2 + 96T^2_1 - 48T_6 + 288T_3 \nonumber \\
& & + 480 T_2 + 1536 T_1 + 2304 \nonumber \\
g^{-1}_{33} &=&  4T_2T^2_1 + 12T^3_1 - 4T_2T_3 - 3T^2_3 + T_6T_1 - 36T_3T_1 \nonumber \\
& & - 60T_1T_2 - 96T^2_1 + 12T_6 - 48T_3 - 48T_2 - 384T_1 \nonumber \\
g^{-1}_{36} &=& 2T_1T_2T_3 - 16T^2_1T_2 - 4T_3T_6 - 6T_2T_6 - 8T_1T_6 + 72T^2_2  \nonumber \\
& &  + 32T_2T_3 - 12T_1T_3 + 144T_1T_2 + 96T^2_1 + 36T_6 \nonumber \\
& & - 240T_3 - 48T_2 - 960T_1 - 2304 \nonumber \\
g^{-1}_{66} &=&  -16T^3_1T_2 - 4T_1T_2T_6 - 8T^2_1T_6 + 2T_2T^2_3 \nonumber \\
& &  + 48T_1T_2T_3 - 96T^2_1T_2 + 48T_1T^2_2 - 192T^3_1 \nonumber \\
& &  -6T^2_6 + 16T_3T_6 - 24T_2T_6 + 16T_1T_6 + 24T^2_3 \nonumber \\
& & + 512 T_2T_3 + 576T_1T_3 + 864T^2_2 + 2880T_1T_2 + 1344T^2_1 \nonumber \\
& & + 96T_6 + 1152T_3 + 6144T_2 + 7680T_1 + 9216
\label{3.2}
\end{eqnarray}
Since $F_4$ has two orbits in the roots we have according to the factorization theorem
\begin{equation}
\det g^{-1} = \frac14 P_1 \cdot P_2
\label{3.3}
\end{equation}
where $P_i$ corresponds to $\Omega_i$ (\ref{2.10}, \ref{2.11}).
We find explicitely
\begin{eqnarray}
P_1 & = & 110592 T_1 + 41472 T_2 + 27648 T_3 + 110592 - 3456 T_6 - 1728 T_1 T_6 \nonumber \\
   & &  + 192 T_1 T_3^2  - 432 T_3 T_6 - 112 T_1^3  T_3 - 384 T_1^2  T_3 \nonumber \\
  & &    + 5184 T_2 T_3 + 20736 T_1 T_2 - 1728 T_1^2  T_2 + 144 T_1^2  T_6 \nonumber \\
  & &    - 48 T_1^3  T_2 - 648 T_2 T_6 + 14976 T_1 T_3 + 4 T_6 T_1^3 \nonumber \\
 & &     - T_3^2  T_1^2  + 216 T_1 T_2 T_3 - 18 T_6 T_1 T_3 - 4608 T_1^3 \nonumber \\
 & &     + 1728 T_3^2  + 18432 T_1^2  + 3888 T_2^2  + 27 T_6^2  + 16 T_1^5 \nonumber \\
 & &     + 4 T_3^3 \label{3.4a} \\
P_2 & = & 10616832 T_1 + 4423680 T_2 + 1769472 T_3 + 7077888 - 221184 T_6 \nonumber \\
  & &    - 221184 T_1 T_6 + 103680 T_1 T_3^2  - 27648 T_3 T_6 \nonumber \\
  & &    + 34560 T_1^3  T_3 + 663552 T_1^2  T_3 + 774144 T_2 T_3 \nonumber \\
  & &    + 4866048 T_1 T_2 + 1465344 T_1^2  T_2 - 62208 T_1^2 T_6 \nonumber \\ 
  & &    - 6912 T_1^3  T_2 - 78336 T_2 T_6 + 1990656 T_1 T_3 \nonumber \\
  & &    - 1728 T_6 T_1^3  + 18144 T_3^2  T_1^2  - T_2^2  T_6^2 \nonumber \\
  & &    + 566784 T_1 T_2 T_3 - 17280 T_6 T_1 T_3 - 8 T_6 T_1 T_2^3 \nonumber \\
  & &    + 79488 T_1^2  T_3 T_2 - 48384 T_6 T_1 T_2 - 2592 T_6 T_1 T_2^2 \nonumber \\
  & &    + 43200 T_1 T_2^2  T_3 + 36 T_6^2  T_1 T_2 - 2592 T_3 T_1^2  T_6 \nonumber \\
  & &    - 8640 T_3 T_1^3  T_2 - 4608 T_2 T_3 T_6 - 5184 T_2 T_1^2  T_6 \nonumber \\
  & &    + 576 T_6 T_1^3  T_2 - 18 T_6 T_2 T_3^2  + 13392 T_1 T_2 T_3^2 \nonumber \\ 
  & &    - 144 T_2^2  T_3 T_6 - 1728 T_2^2  T_3 T_1^2  + 96 T_2^3  T_3 T_1 \nonumber \\
  & &    - 216 T_3^2  T_1^2  T_2 - 108 T_1 T_6 T_3^2  + 72 T_1^2  T_6 T_2^2 \nonumber \\
  & &    + 774144 T_1^3  + 138240 T_3^2  + 5308416 T_1^2  + 1096704 T_2^2 \nonumber \\
  & &    + 1728 T_6^2  - 20736 T_1^5  + 3456 T_3^3  - 103680 T_1^4  + 129024 T_2^3 \nonumber \\
  & &    + 6384 T_2^4  + 1728 T_1^6  + 27 T_3^4  + 64 T_2^5  + 4 T_6^3 \nonumber \\
  & &    + 119808 T_3 T_2^2  + 787968 T_1 T_2^2  + 36288 T_2 T_3^2 \nonumber \\
  & &    + 88128 T_1  T_2  + 108 T_1  T_6  + 45888 T_1 T_2 \nonumber \\
  & &    - 18432 T_1^3  T_2^2  - 9024 T_2^2  T_6 + 2520 T_2^2  T_3^2 \nonumber \\
  & &    + 192 T_2 T_6^2  + 864 T_1 T_6^2  - 432 T_6 T_3^2  - 328 T_2^3  T_6  \nonumber \\
  & &    - 10368 T_1^4  T_3 - 32 T_1^3  T_2^3  + 6592 T_2^3  T_3 - 2976 T_1^2  T_2^3 \nonumber \\
  & &    + 432 T_1^4  T_2^2  + 1728 T_1^5  T_2 - 432 T_3^2  T_1^3  + 1296 T_3^3  T_1 \nonumber \\
  & &    + 864 T_1^4  T_6 + 432 T_2 T_3^3  - 34560 T_1^4  T_2 + 32 T_2^4  T_3 \nonumber \\
  & &    - 16 T_1^2  T_2^4  + 224 T_1 T_2^4  + 4 T_2^3  T_3^2 \nonumber \\
  & &    - 1296 T_6 T_1 T_2 T_3 \label{3.5a} 
\end{eqnarray}
If we reexpress these polynomials in the Chevalley basis $T_1,T_2,T_3,T_6$ as functions of 
Cartesian coordinates we find
\begin{eqnarray}
P_1 & = & - 2^{24} \prod_{\alpha \in \Omega_1^+} [\sin{\frac12 (\alpha,x)}]^2 \label{3.6} \\
P_2 & = & - 2^{24} \prod_{\alpha \in \Omega_2^+} [\sin{\frac12 (\alpha,x)}]^2 \label{3.7} 
\end{eqnarray}

Here we made use of the fact that each orbit $\Omega_{k,a}$ can be decomposed into a positive and 
a negative semiorbit
\begin{eqnarray}
\Omega_{k,a}=\Omega_{k,a}^+ \cup \Omega_{k,a}^-
\end{eqnarray}
by a hypersurface 
\begin{eqnarray}
(\alpha,\xi)=0
\end{eqnarray}
so that
\begin{eqnarray} 
\Omega_{k,a}^{\pm}=\{ \alpha \in \Omega_{k,a}; (\alpha,\xi) \gtrless 0 \}
\end{eqnarray}
For example, we define
\begin{eqnarray}
\Omega_{k,a}^+ & = & \{\mu_1 e_1 +\mu_2 e_2 +\mu_3 e_3 + \mu_4 e_4 \in \Omega_{k,a}; \\
               &  & \mu_1 > 0 \textrm{ or } \mu_1=0 , \mu_2 > 0  \\ 
               &  & \textrm{ or } \mu_1=\mu_2=0, \mu_3>0 \\
               &  & \textrm{ or } \mu_1=\mu_2=\mu_3=0 , \mu_4 >0 \}
\end{eqnarray}
Then the coefficients $\xi_n$ of $\xi$ 
\begin{eqnarray}
\xi=\sum_n \xi_n e_n
\end{eqnarray}
must satisfy inequalities, i.e. from $\Omega_1$ and $\Omega_2$ 
\begin{eqnarray}
\xi_1 > \xi_2 > \xi_3 > \xi_4 > 0 \label{3.12a} \\
\xi_1 > \xi_2 + \xi_3 + \xi_4 \label{3.12b}
\end{eqnarray}
$\Omega_3$ implies new inequalities, etc., so that a vector $\xi$ exists for any finite set of orbits. 
We assume (\ref{3.12a}),(\ref{3.12b}) to hold throughout. 

We consider the asymptotic behaviour along an imaginary direction in $x$--space
\begin{eqnarray}
x= -i \lambda \xi , \quad \lambda \rightarrow \infty \label{3.14}
\end{eqnarray}
In each orbit there exist maximal vectors $\alpha_{\max}(k,a)$ so that for fixed $\xi$ 
\begin{eqnarray}
(\alpha_{\max} (k,a),\xi)= \max_{\alpha \in \Omega_{k,a}}(\alpha,\xi) \label{3.14a} 
\end{eqnarray} 
Then (for unique $\alpha_{\max}(k,a)$)
\begin{eqnarray}
T_{k,a} \sim e^{\lambda(\alpha_{\max}(k,a),\xi)} \label{3.15}\\
\textrm{ for } \lambda \rightarrow \infty
\end{eqnarray}
In particular, we have 
\begin{eqnarray}
\alpha_{\max}(1) & = & e_1 \\
\alpha_{\max}(2) & = & e_1+e_2 \\
\alpha_{\max}(3) & = & \frac12 (3e_1+e_2+e_3+e_4) \\
\alpha_{\max}(4) & = & 2e_1+e_2+e_3
\end{eqnarray}
From (\ref{3.14}),(\ref{3.14a}) we find as leading term in $P_1(P_2)$ the unique monomial 
$-T_1^2T_3^2(-T_2^2T_6^2)$ with asymptotic behaviour 
\begin{eqnarray}
P_{1,2} \sim -e^{\lambda (\rho_{1,2},\xi)} \label{3.17}
\end{eqnarray}
where 
\begin{eqnarray}
\rho_1 & = & 5e_1+e_2+e_3+e_4 \nonumber \\
\rho_2 & = & 6e_1+4e_2+2e_3 \label{3.18}
\end{eqnarray}
These are special cases of Weyl type vectors 
\begin{eqnarray}
\rho_{k,a}=\sum_{\alpha \in \Omega_{k,a}^+} \alpha \label{3.19}
\end{eqnarray}
The same leading term (\ref{3.17}) results from (\ref{3.6}), (\ref{3.7}). In this fashion the constants 
in the factorization formulas (\ref{3.6}), (\ref{3.7}) can be controlled.

From (\ref{2.23}) follows that 
\begin{eqnarray}
\textrm{det } g^{-1}=\frac14 P_1 P_2 =(\textrm{det }J)^2 \label{3.20} 
\end{eqnarray}
where $J$ is the Jacobian matrix (\ref{2.22}). Due to (\ref{3.6}),(\ref{3.7}) det$J$ vanishes 
in first order along the hyperplanes 
\begin{eqnarray}
(\alpha,x)=2n\pi, \quad n\in \Bbb Z, \alpha \in \Omega_{1,2} \label{3.21}
\end{eqnarray}
The root space ${\Bbb R}_4$ is therefore divided into cells bounded by the walls (\ref{3.21}) 
where (see (\ref{3.32})) repulsive and impenetrable potentials are positioned. 

Next we derive the $r$--coefficients from 
\begin{eqnarray}
r_m^{(a)}=\sum_n g_{mn}^{-1} \frac{\partial \log P_a}{\partial T_n} \label{3.22}
\end{eqnarray}
that ought to be polynomials in the $\{T_n\}_{n \in \{1,2,3,6\}}$. In fact, we find 
\begin{eqnarray}
r^{(1)} & = & (-5 T_1-24,-6T_2-6T_1, -9T_3-12T_2-24T_1,-4T_1T_2-12T_6-24T_2+24T_1) \nonumber \\
 \label{3.23} \\
r^{(2)} & = & (-6T_1,-10T_2-48,-12T_3-24T_1,-24T_1^2-18T_6+48T_3+96T_2+192T_1+576) \nonumber \\ 
\label{3.24}
\end{eqnarray}
From (\ref{3.2}) and (\ref{3.23}),(\ref{3.24}) we obtain the algebraic differential operator 
\begin{eqnarray}
D=-\sum_{m,n} \frac{\partial}{\partial T_m} g_{mn}^{-1} \frac{\partial}{\partial T_n} +\sum_{a=1,2} \gamma_a \sum_m r_m^{(a)} \frac{\partial}{\partial T_m} \label{3.25}
\end{eqnarray}
containing two real coupling constants $\gamma_a$ as free parameters. This operator is exactly 
solvable in terms of polynomial eigenfunctions
\begin{eqnarray}
D p_{\lambda}(T_1,T_2,T_3,T_6)= \lambda p_{\lambda}(T_1,T_2,T_3,T_6) \label{3.26}
\end{eqnarray}
One can show that $D$ possesses a flag of invariant polynomial spaces $\{V_N\}$ 
\begin{eqnarray}
& & D V_N \subset V_N \label{3.27} \\
V_N & = & \textrm{span} \{T_1^{n_1}T_2^{n_2}T_3^{n_3}T_6^{n_6}, n_i \in {\Bbb Z}_{\geq}, \\
    & & n_1+2n_2+3n_3+4n_6 \leq N\} \label{3.28}
\end{eqnarray}
Thus the eigenfunctions $\{p_{\lambda}\}$ can be calculated by linear algebra. 

On the other hand we know from the general scheme that $D$ corresponds by a ''gauge transformation'' \cite{1} 
to the Schr\"odinger operator in ${\Bbb R}_4$ 
\begin{eqnarray}
{\cal{H}}=-\Delta + W \label{3.29}
\end{eqnarray}
with a Laplacian $\Delta$ and a potential $W$
\begin{eqnarray}
W & = & \frac14 \sum_{a,b} (\gamma_{a}\gamma_{b}-\frac14) R_{ab} \label{3.30} \\
R_{ab}& = & \sum_{m,n} g_{mn}^{-1} \frac{\partial \log P_a}{\partial T_m} \frac{\partial \log P_b}{\partial T_n} \nonumber \\
& = & \sum _n r_n^{(a)} \frac{\partial}{\partial T_n} \log P_b \label{3.31}
\end{eqnarray}
In Cartesian coordinates we can for $F_4$ express
\begin{eqnarray}
R_{ab}=\rho_{ab}\sum_{\alpha \in \Omega_a^{+}} [\sin{\frac12 (\alpha,x)}]^{-2} +C_{ab} \label{3.32}
\end{eqnarray}
where 
\begin{eqnarray}
\begin{array}{cc}
\rho_{11}=+1, & C_{11}=-28 \label{3.33} \\
\rho_{22}=+2, & C_{22}=-56 \label{3.34} \\ 
\rho_{12}=\rho_{21}=0, & C_{12}=C_{21}=-36 \label{3.35}
\end{array}
\end{eqnarray}
can be determined by an asymptotic analysis.

In this fashion we have derived the Sutherland model for $F_4$.

\vspace{1,5cm}

\appendix
\section{Tables}

{\bf{Table 1}}: Orbits, generating vectors and cardinals for $F_4$

\begin{eqnarray*}
\begin{array}{llr}
\Omega \qquad\quad & \sum n_ie_i  & \quad \#\Omega \\ \hline
\Omega_1 & e_1 & 24 \\
\Omega_2 & e_1 + e_2 & 24 \\
\Omega_3 & e_1 + e_2 + e_3 & 96 \\
\Omega_4 & 2e_1, e_1 + e_2 + e_3 + e_4 & 24 \\
\Omega_5 & 2e_1 + e_2 & 144 \\
\Omega_6 & 2e_1 + e_2 + e_3 & 96 \\
\Omega_7 & 2e_1 + e_2 + e_3 + e_4 & 192 \\
\Omega_8 & 2e_1 + 2e_2 & 24 \\
\Omega_{9,1} & 2e_1 + 2e_2 + e_3 & 288 \\
\Omega_{9,2} & 3e_1 & 24 \\
\Omega_{10} & 3e_1 + e _2, \; 2e_1 + 2e_2 + e_3 + e_4 & 144 \\
\Omega_{11} & 3e_1 + e_2 + e_3 & 288 \\
\Omega_{12} & 2e_1 + 2e_2 + 2e_3, \; 3e_1 + e_2 + e_3 + e_4 & 96 \\ 
\Omega_{13,1} & 3e_1 + 2e_2 & 144 \\
\Omega_{13,2} & 2e_1 + 2e_2 + 2e_3 + e_4 & 192 \\
\Omega_{14} & 3e_1 + 2e_2 + e_3 & 192 \\
\Omega_{15} & 3e_1 + 2e_2 + e_3 + e_4 & 576 \\
\Omega_{16} & 2e_1 + 2e_2 + 2e_3 + 2e_4, \; 4e_1 & 24 \\
\Omega_{17,1} & 3e_1 + 2e_2 + 2e_3 & 288 \\
\Omega_{17,2} & 4e_1 + e_2 & 144 \\
\Omega_{18,1} & 3e_1 + 3e_2 & 24 \\
\Omega_{18,2} & 4e_1 + e_2 + e_3, \; 3e_1 + 2e_2 + 2e_3 + e_4 & 288 \\
\Omega_{19,1} & 3e_1 + 3e_2 + e_3 & 288 \\
\Omega_{19,2} & 4e_1 + e_2 + e_3 + e_4 & 192 \\
\Omega_{20} & 4e_1 + 2e_2, \; 3e_1 + 3e _2 + e_3 + e_4 & 144 \\
\Omega_{21,1} & 4e_1 +2e_2 + e_3 & 576 \\
\Omega_{21,2} & 3e_1 + 2e_2 + 2e_3 + 2e_4 & 192 \\
\Omega_{22} & 3e_1 + 3e_2 +2 e_3, \; 4e_1 + 2e_2 + e_3 + e_4 & 288 \\ 
\Omega_{23} & 3e_1 + 3e_2 + 2e_3 + e_4 & 576 \\
\Omega_{24} & 4e_1 + 2e_2 + 2e_3 & 96
\end{array} 
\end{eqnarray*}

{\bf{Table 2}}: $T_{n,a}$ expressed as polynomial in $T_1,T_2,T_3,T_6$ for $F_4$
\begin{eqnarray*}
T_4 &=& T^2_1 - 2T_3 - 6T_2 - 8T_1 - 24 \\
T_5 &=& T_2T_1 - 3T_3 - 6T_1 \\
T_7 &=& T_3T_1 - 4T_2T_1 - 8T^2_1 - 3T_6 + 22T_3 + 36T_2 + 80T_1 + 192 \\
T_8 &=& T^2_2 - 6T^2_1 - 2T_6 + 12T_3 + 28T_2 + 48T_1 + 120 \\
T_{9,1} &=& T_3T_2 - 3T_3T_1 + 8T_2T_1 + 24T^2_1 + 9T_6 - 60 T_3 - 108T_2 - 228T_1 - 576 \\
T_{9,2} &=& T^3_1 - 3T_3T_1 - 3T_2T_1 + 3T_6 - 21T_3 - 36T_2 - 99T_1 - 192 \\
T_{10} &=& T_2T^2_1 - 2T_3T_2 - 8T_2T_1 - 6T^2_2 - 3T_6 - 30T_2 \\
T_{11} &=& T_6T_1 - 2T_3T_2 + 3T_3T_1 - 8T_2T_1 - 24T^2_1 - 9T_6 + 63T_3 + 108T_2 + 240T_1 + 576 \\
T_{12} &=& -4T_2T^2_1 -8T^3_1 -2T_6T_1 + 8T_3T_2 + 12T^2_2 + 24T_3T_1 + 56T_2T_1 + T^2_3 + 60T^2_1 \\
& & + 40T_3 + 120T_2 + 288T_1 + 288 \\
T_{13,1} &=& T^2_2T_1 - 6T^3_1 - 2T_6T_1 - T_3T_2 + 15T_3T_1 + 19T_2T_1 + 24T^2_1 -9T_6 \\
& & + 63T_3 + 108T_2 + 354T_1 + 576 \\
T_{13,2} &=& T_3T^2_1 - 2T^2_3 - T_6T_1 - 6T_3T_2 - 6T_3T_1 - 8T_2T_1 - 16T^2_1 - 6T_6 \\
& & + 20T_3 + 72T_2 + 152T_1 + 384\\
T_{14} &=& 8T_2T^2_1 + 24T^3_1 + T_6T_2 + 6T_6T_1 - 72T_3T_1 - 136T_2T_1 - 20T^2_2 - 16T_3T_2 - 3T^2_3 \\
& & - 144T^2_1 + 22T_6 - 192T_3 - 400T_2 - 1152T_1 - 1536 \\
T_{15} &=& T_3T_2T_1 - 3T_3T^2_1 - 4T^2_2T_1 - 8T_2T^2_1 - 3T_6T_2 + T_6T_1 + 40T_3T_2 + 21T_3T_1 \\
& & 96T_2T_1 + 6T^2_3 + 36T^2_2 +24T_1^2+ 9T_6-6T_3 + 84T_2 - 240T_1 - 576 \\
T_{16} &=& T^4_1 - 4T_3T^2_1 - 4T_2T^2_1 + 4T_6T_1 + 2T^2_3 + 8T_3T_2 + 6T^2_2 - 16T_3T_1 \\
& &- 16T_2T_1 - 76T^2_1 + 12 T_6 - 40T_3 - 72T_2 - 416T_1 - 552 \\
T_{17,1} &=& -2T_3T_2T_1 + 3T_3T^2_1 + 4T^2_2T_1 + 16T_2T^2_1 + 12T^3_1 + T_6T_3 + 6T_6T_2 + 5T_6T_1 \\
& & -6T^2_3 - 55T_3T_2 - 72T^2_2 - 51T_3T_1 - 200T_2T_1 - 72T^2_1 + 9T_6 - 120 T_3 \\
& & - 492T_2 - 468T_1 - 576 \\
T_{17,2} &=& -3T_3T_2T_1 + T_2T^3_1 - 3T_2^2T_1 + 3T_6T_2 - T_6T_1 - 36T^2_2 - 19T_3T_2 \\
& & - 3T_3T_1 - 92T_2T_1 + 24T^2_1 + 9T_6 - 60T_3 - 300T_2 - 234T_1 - 576 \\
T_{18,1} &=& T^3_2 - 24T^3_1 - 15T_2T^2_1 - 3T_6T_2 - 6T_6T_1 + 3T^2_3 + 30T_3T_2 + 72T_3T_1 + 54T^2_2 \\
& & + 192T_2T_1 + 144T^2_1 - 21T_6 + 192T_3 + 549T_2 + 1152T_1 + 1536 \\
T_{18,2} &=& T_6T^2_1 - 28T_2T^2_1 - 72T^3_1 - 2T_6T_3 - 9T_6T_2 - 26T_6T_1 + 56T_3T_2 + 216T_3T_1 + 84T^2_2  \\
& &+ 440 T_2T_1 + 9T^2_3 + 432T^2_1 - 84T_6 + 576T_3 + 1308T_2 + 3456T_1 + 4608 \\
T_{19,1} &=& T_3T^2_2 - T_3T_2T_1 + 8T_2T^2_1 + 4T^2_2T_1 - 3T_3T^2_1 - 2T_6T_3 + 3T_6T_2 - 3T_6T_1 + 6T^2_3 \\
& & - 8T_3T_2 - 36T^2_2 + 18T_3T_1 - 68T_2T_1 + 48T^2_1 + 18T_6 - 69T_3 - 408T_2 \\
& &- 480T_1 - 1152 \\
T_{19,2} &=& -8T^4_1 - 4T_2T^3_1 + T^2_3T_1 + 9T_3T_2T_1 + 23T_3T^2_1 + 12T^2_2T_1 - 2T_6T^2_1 + 48T_2T^2_1 \\
& & + 48 T^3_1 - T_6T_3 - 3T_6T_2 - 7T_6T_1 + 2T^2_3 + 24T_3T_2 + 36T^2_2 + 72T_3T_1 + 244T_2T_1 \\
& & + 384T^2_1 + 18T_3 + 192T_2 + 512T_1 \\ 
T_{20} &= & -1716 T_2 + 48 T_6 - 2 T_3 T_2^2  - 8 T_1 T_2^2  - 2 T_1^2  T_6 - 2928 T_1 - 636 T_3 \\
& & + 76 T_1^2  T_2 + T_1^2  T_2^2  - 228 T_2^2 - 152 T_2 T_3 + 22 T_1 T_6 - 264 T_1 T_3 - 306 T_1^2 \\
& &  - 680 T_1 T_2 + 12 T_2 T_6 - 27 T_3^2  + 24 T_1^2  T_3 + 120 T_1^3 + 4 T_3 T_6 - 6 T_2^3  - 3600 - 6 T_1^4 
\end{eqnarray*}
\begin{eqnarray*}
T_{22} & = & -4608-1008T_2+111T_6-576T_3-3456T_1+34T_1T_6 
     + 4 T_3 T_6 + 8 T_3 T_2^2  - 16 T_2 T_3  \\
   & & - 216 T_1 T_3 - 152 T_1 T_2 + 88 T_1^2  T_2 + 56 T_1 T_2^2  - 2 T_1^2  T_6  
     - 8 T_1^3  T_2 + 15 T_2 T_6 + T_2 T_3^2  \\
   & & - 4 T_1^2  T_2^2 - 2 T_6 T_1 T_2 + 24 T_1 T_2 T_3 - 432 T_1^2  - 9 T_3^2  + 36 T_2^2 
    + 72 T_1^3  + 12 T_2^3 \\
T_{24} & = & 4320 + 1104 T_2 - 152 T_6 + 912 T_3 + 3648 T_1 - 64 T_1 T_6 \\
     & & - 16 T_3 T_6 - 8 T_3 T_2^2  + 96 T_2 T_3 + 384 T_1 T_3 \\
     & & + 128 T_1 T_2 - 208 T_1^2  T_2 - 80 T_1 T_2^2  + 8 T_1^2  T_6 \\
     & & + 16 T_1^3  T_2 - 24 T_2 T_6 - 2 T_2 T_3^2  + 4 T_1^2  T_2^2 \\
    & & - 48 T_1^2  T_3 + 4 T_6 T_1 T_2 - 48 T_1 T_2 T_3 + 312 T_1^2 \\
    & & + 48 T_3^2  + 12 T_2^2  - 192 T_1^3  - 8 T_2^3  + 12 T_1^4  + T_6^2 
\end{eqnarray*}

\end{document}